\documentclass{article}
\usepackage{spconf,amsmath,amssymb,graphicx,hyperref,mathtools,subcaption,bbm}

\usepackage[dvipsnames]{xcolor}
\usepackage[most]{tcolorbox}

\usepackage{multirow}
\usepackage{listings}
\newcommand{\ignore}[1]{}

\definecolor{codegreen}{rgb}{0,0.6,0}
\definecolor{codegray}{rgb}{0.5,0.5,0.5}
\definecolor{codepurple}{rgb}{0.58,0,0.82}
\definecolor{backcolour}{rgb}{0.95,0.95,0.92}

\lstdefinestyle{mystyle}{
    backgroundcolor=\color{backcolour},   
    commentstyle=\color{codegreen},
    keywordstyle=\color{codepurple},
    numberstyle=\tiny\color{codegray},
    stringstyle=\color{codepurple},
    basicstyle=\ttfamily\footnotesize,
    breakatwhitespace=false,         
    breaklines=true,                 
    captionpos=b,                    
    keepspaces=true,                                  
    numbersep=5pt,      
    frame = shadowbox,
    showspaces=false,                
    showstringspaces=false,
    showtabs=false,                  
    tabsize=2
}

\newcommand{\boxedthm}[1]{
\begin{tcolorbox}[colback=gray!30,
                  colframe=black,
                  width=\linewidth,
                  arc=2mm, auto outer arc,
                  boxrule=1pt,
                  boxsep=-1mm,
                 ]
  #1
\end{tcolorbox}
}

\usepackage{amsthm}

\newtheoremstyle{withdot}
  {\topsep}   
  {\topsep}   
  {\itshape}  
  {}          
  {\bfseries} 
  {.}         
  {.5em}      
  {}          

\newtheorem{theorem}{Theorem}

\usepackage{xcolor}
\newcommand{\ind}[1]{\mathbbm{1}_{\{#1\}}}
\newcommand{\subcref}[2]{%
  Fig.~\hyperref[#1]{\getrefnumber{#1}(#2)}%
}

\title{Real-Time Markov Modeling for Single-Photon LiDAR: \\ 1000× Acceleration and Convergence Analysis}
\name{Weijian Zhang, Hashan K. Weerasooriya, Prateek Chennuri, Stanley H. Chan
\thanks{The work is supported, in part, by the DARPA / SRC CogniSense JUMP 2.0 Center, NSF IIS-2133032, and NSF ECCS-2030570.}}
\address{School of Electrical and Computer Engineering, Purdue University, West Lafayette, USA}
\begin{document}
\ninept
\maketitle
\begin{abstract}
Asynchronous single-photon LiDAR (SP-LiDAR) is an important imaging modality for high-quality 3D applications and navigation, but the modeling of the timestamp distributions of a SP-LiDAR in the presence of dead time remains a very challenging open problem. Prior works have shown that timestamps form a discrete-time Markov chain, whose stationary distribution can be computed as the leading left eigenvector of a large transition matrix. However, constructing this matrix is known to be computationally expensive because of the coupling between states and the dead time. This paper presents the first non-sequential Markov modeling for the timestamp distribution. The key innovation is an equivalent formulation that reparameterizes the integral bounds and separates the effect of dead time as a deterministic row permutation of a base matrix. This decoupling enables efficient vectorized matrix construction, yielding up to $1000 \times$ acceleration over existing methods. The new model produces a nearly exact stationary distribution when compared with the gold standard Monte Carlo simulations, yet using a fraction of the time. In addition, a new theoretical analysis reveals the impact of the magnitude and phase of the second-largest eigenvalue, which are overlooked in the literature but are critical to the convergence. 
\end{abstract}
\section{Introduction}
\label{sec:intro}

\begin{figure}[t]
    \centering
    \includegraphics[width=\linewidth]{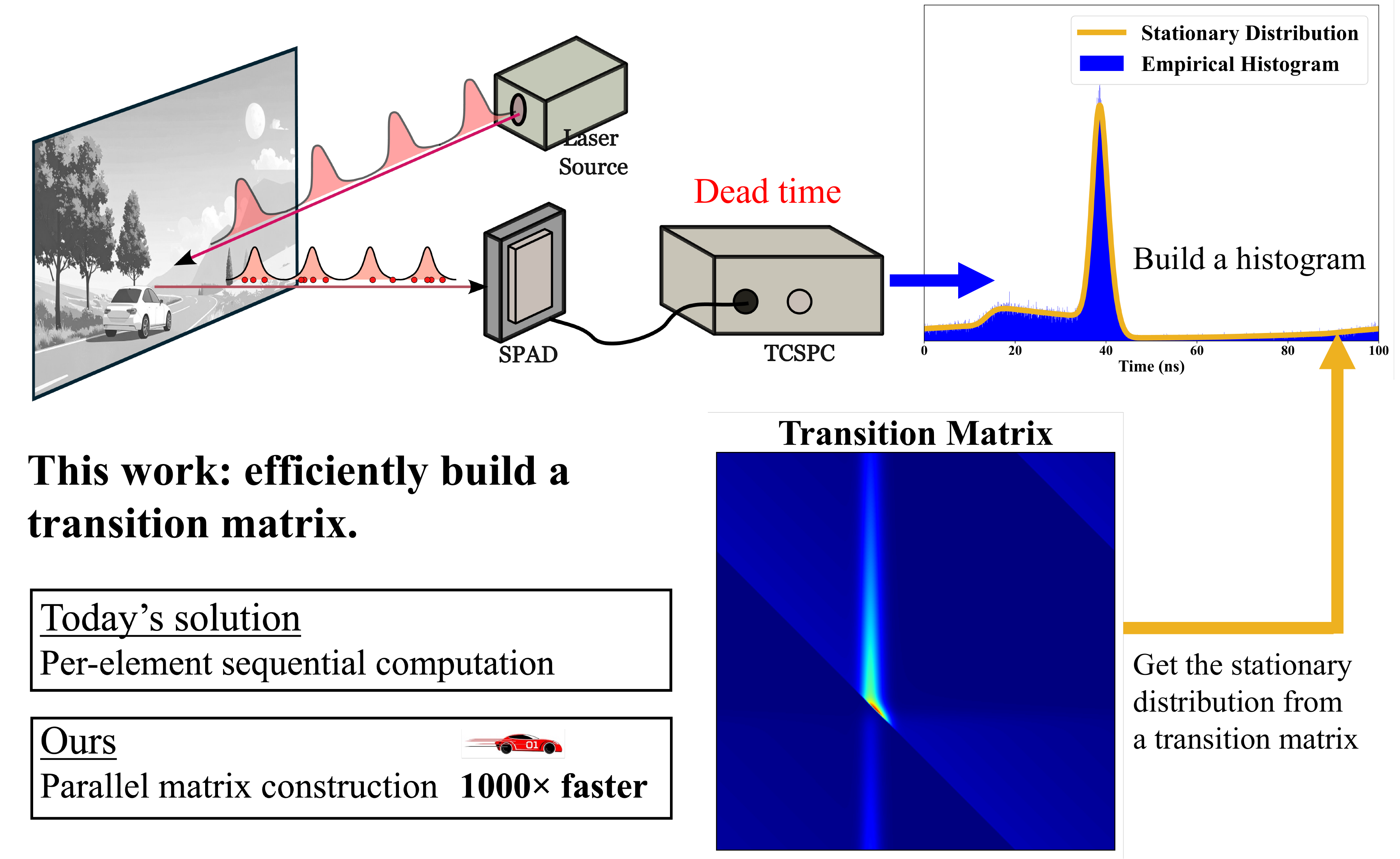}
    \caption{\textbf{This paper addresses modeling of timestamp distributions in asynchronous SP-LiDAR.} We show that they can be rapidly obtained by deriving a parallelizable transition matrix.}
    \label{fig:lidar_deadtime}
    \vspace{-17pt}
\end{figure}

Single-photon LiDAR (SP-LiDAR) is increasingly used in applications requiring ultra-low-light sensitivity and picosecond timing resolution \cite{Li_2020_LiDARReview, mora_martin_21_object_detection, McCarthy_2025_long_distance, goyal_2025_robust3dobjectdetection, tobin_2021_obscurant, Scheuble_2025_waveforms, malik_2025_inverse_rendering, kavinga_2025_joint_estimation}. In an asynchronous operation mode, the system enforces a fixed \emph{dead time} after each detection. During this period, the detector is inactive and becomes rearmed after the dead time ends. Since there is no synchronization with the laser, the stochastic reactivation times result in a highly nonlinear and scene-dependent distortion of the observed timestamp distribution, as shown in Fig.~\ref{fig:lidar_deadtime}.

To model this behavior, recent works have shown that the sequence of relative timestamps can be described by a Markov chain whose transition probabilities capture the combined effects of signal flux, background flux, and dead time~\cite{Rapp_2019_Dead,rapp_2021_high}. The stationary distribution of this chain corresponds to the long-term probability density function (PDF) of photon timestamps from which we can estimate the depth, model the histogram, and simulate the system.

Despite its theoretical appeal, the Markov chain approach suffers from a severe computational bottleneck: constructing the transition matrix involves evaluating complex integrals that depend jointly on the current state, the next state, and the dead time duration. Consequently, constructing the matrix becomes computationally expensive and difficult to parallelize, particularly at high temporal resolutions where the number of states is large.

In this work, we introduce a new formulation of the transition probability matrix by decoupling the effect of dead time from the rest of the system. We derive an equivalent representation of the integral bounds that eliminates the dead time dependency from the base matrix and show that dead time acts as a deterministic row-wise shift applied to the base matrix. This new observation enables the entire matrix to be constructed in a vectorized and parallelizable fashion, leading to up to \textbf{1000× acceleration} compared to the baseline method.

Because of the computational gain, our new model unlocks new capabilities for large-scale analysis. For various signal and background levels, we evaluate the accuracy of stationary PDFs against Monte Carlo simulations and analyze the convergence properties of the chains. Specifically, we observe, for the first time, that the \textbf{phase} of the second-largest eigenvalue governs the oscillatory component of the convergence behavior, something that has often been overlooked in spectral mixing analysis.

Our contributions are summarized as follows:
\begin{itemize}
    \item We introduce a new representation of the transition matrix for asynchronous SP-LiDAR. Our new model separates dead time by treating it as a permutation operator.
    \item We propose a vectorized matrix construction algorithm that scales efficiently with the temporal resolution, enabling up to $1000 \times$ speedup over previous methods.
    \item Leveraging this efficient formulation, we validate the predicted stationary distributions across diverse flux conditions and analyze the chain's spectral convergence dynamics — analyses previously impractical due to computational cost.
\end{itemize}

\section{Background}
\label{sec:bg}

Single-photon LiDAR (SP-LiDAR) is different from conventional linear LiDAR in terms of sensitivity, timing resolution, and cost. In SP-LiDAR, we are more interested in individual timestamps, while in conventional LiDAR, the subject of interest is the point cloud.

SP-LiDAR records photon arrival times using a single-photon avalanche diode (SPAD) detector and a time-correlated single-photon counting (TCSPC) module. Both components exhibit dead time after each detection or registration, during which subsequent photons are lost. At high flux, this effect introduces nonlinear distortions in the measured histogram.

\subsection{Photon Arrival and Detection Model}

Following assumptions from prior works~\cite{Shin_2015_3D,rapp_2017_unmixing,Chan_2024_CVPR}, we model photon arrivals as an inhomogeneous Poisson process~\cite{Bar-David_1969,Snyder_1991_book}, with rate
\begin{equation}
    \label{eq:arrival_flux}
    \lambda(t) = \alpha\, s(t-\tau) + \lambda_b,
    \quad t \in [0,t_r),
\end{equation}
known as the photon arrival flux function (one period), where $\alpha$ is the surface reflectivity, $\tau$ is the depth-induced delay, $\lambda_b$ is the constant background illumination, and $t_r$ is the laser repetition period. The transmitted laser pulse is modeled as $s(t) = \calE \calN(t;0,\sigma_t^2)$, a Gaussian-shaped function with pulse energy $\calE$ and width $\sigma_t$. Defining the signal level $S \coloneq \alpha \calE$, noise level $B \coloneq t_r \lambda_b$, and the total flux $\Lambda = S+B$, the normalized arrival distribution is
\begin{equation}
    f_a(t) = \frac{\lambda(t)}{\Lambda} = \frac{S}{\Lambda}\,\calN(t-\tau) + \frac{B}{\Lambda}\,\frac{1}{t_r}, 
    \quad t \in [0,t_r).
    \label{eq:arrival_pdf}
\end{equation}

We consider asynchronous (free-running) detection with SPAD dead time $t_d$, which exceeds the TCSPC dead time so that only the SPAD effect is relevant~\cite{isbaner_2016_dead}. In this mode, the detector reactivates the moment the dead time ends without waiting for the laser sync, enabling multiple photon registrations per cycle. Let $\{T_k\}_{k \in \N}$ denote the absolute detection times; the TCSPC module records relative timestamps $\{X_k\}_{k \in \N}$ via $X_k = T_k \bmod t_r$. A histogram over $[0,t_r)$ with $n_b$ bins is then formed from $\{X_k\}_{k \in \N}$.

\subsection{Distortion of Timestamp Distribution}

Understanding the timestamp distribution is essential, as it serves as the statistical model of the measured histogram and enables principled inference of scene parameters such as depth. When $t_d=0$, all arriving photons are registered, so they are independent and identically distributed (i.i.d.) according to $f_a(t)$~\cite{Bar-David_1969,vivek_2024_detection}.

For $t_d>0$, however, $\{X_k\}_{k \in \N}$ are dependent, and the resulting histogram deviates from \eqref{eq:arrival_pdf}. The distortion depends on $\boldsymbol{\theta} = (t_r,t_d,\sigma_t,\tau,S,B, n_b)$ and includes effects such as peak shifts and noise aggregations~\cite{Rapp_2019_Dead,zhang_2024_mmsp,zhang_2025_icip}. A closed-form registration distribution $f_r(t; \boldsymbol{\theta})$ is generally unavailable, motivating the need for accurate and efficient modeling.

\subsection{Related Work and Preliminaries}
\label{ssec:related_work}
Unlike dead time studies on synchronous SP-LiDAR \cite{coates_1967_pile_up_correction,walker_2002_iterative_pileup,Pediredla_2022_adaptive_gating}, where the timestamp statistic is given by the multinomial distribution~\cite{heide_2018_sub,Pediredla_2018_pileup,Gupta_2019_Asynchronous}, only limited work has explicitly modeled the registration distribution $f_r(t)$ in asynchronous SP-LiDAR systems.

Isbaner \emph{et al.} proposed an iterative estimation method for forward distortion, but its accuracy degraded under high flux~\cite{isbaner_2016_dead}. Zhang \emph{et al.} introduced a Gaussian--uniform mixture model to approximate $f_r(t)$~\cite{zhang_2024_mmsp}, although the mapping from system parameters $\boldsymbol{\theta}$ to mixture parameters remained unclear. More recently, Zhang \emph{et al.} used an autoencoder to learn the distortion~\cite{zhang_2025_icip}; inference was fast, but the network required retraining for each dead time setting. Kitichotkul \emph{et al.} addressed free-running modes~\cite{Kao_2025_freerunning}, but emphasized estimation over modeling of $f_r(t)$.

The most rigorous modeling was developed by Rapp \emph{et al.}~\cite{Rapp_2019_Dead,rapp_2021_high}, who derived a Markov chain representation of the relative timestamps $\{X_k\}_{k \in \N}$ and predicted $f_r(t)$ as the stationary distribution obtained from the leading left eigenvector of a transition matrix. While accurate, their formulation required per-element integration to build the transition matrix, making high-resolution predictions computationally prohibitive. Our work addresses this limitation by deriving equivalent forms that enable parallel matrix construction, thereby accelerating the prediction of $f_r(t)$ by several orders of magnitude.

\noindent\textbf{Markov chain pipeline.}
The registration process can be modeled as a Markov chain where the next timestamp depends on the current one through a transition PDF. This chain is irreducible, recurrent, and aperiodic, ensuring a unique stationary distribution~\cite{Rapp_2019_Dead}.

In practice, the continuous state space $[0,t_r)$ is discretized into $n_b$ TCSPC bins with spacing $\Delta = t_r/n_b$, forming the state set $\mathcal{S} = \{s_1,\dots,s_{n_b}\}$. The transition kernel between states is obtained by discretizing the transition PDF, yielding an unnormalized transition matrix $\mP$. Normalizing each row produces a stochastic matrix $\Tilde{\mP}$. The stationary distribution $\vpi$, satisfying $\vpi \Tilde{\mP} = \vpi$, can then be approximated as the leading left eigenvector of $\Tilde{\mP}$. As shown in~\cite{Rapp_2019_Dead}, this procedure provides an accurate surrogate of the photon registration distribution $f_r(t)$.

However, constructing $\mP$ requires evaluating all $n_b^2$ state pairs. When $n_b$ is large, this step becomes the computational bottleneck, which requires a more efficient formulation.

\section{Accelerated Markov Modeling}

\begin{figure*}[t]
    \centering
    \includegraphics[width=0.96\linewidth]{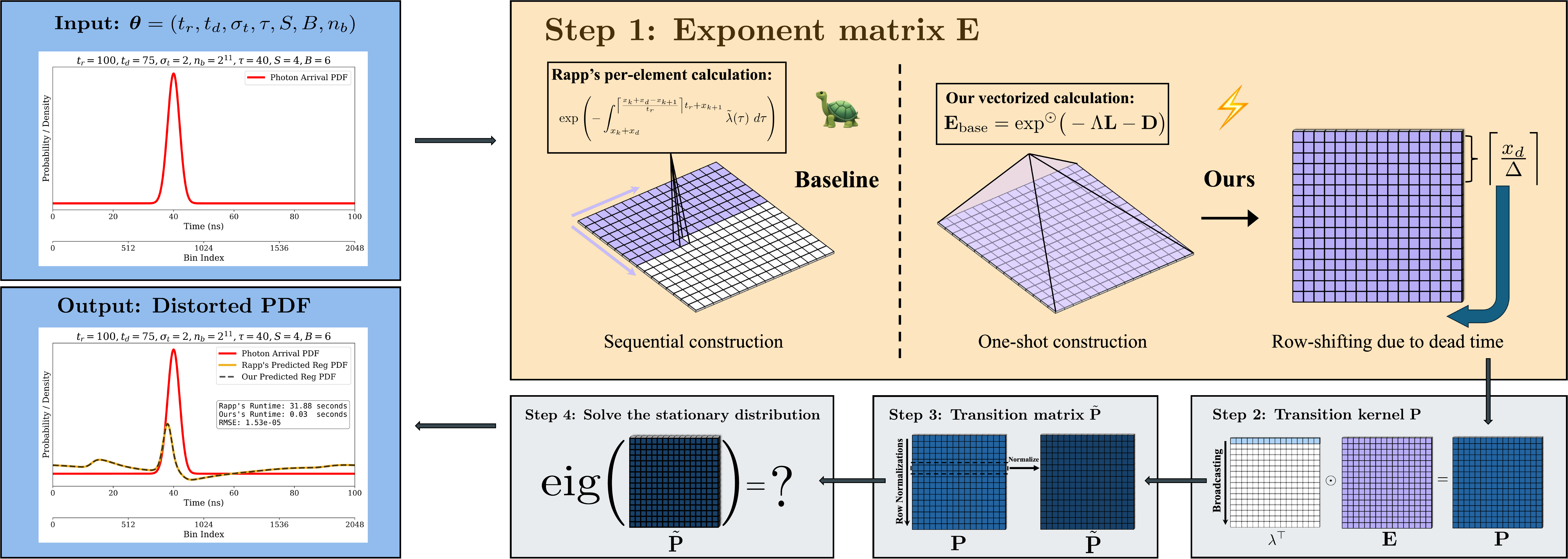}
    \caption{\textbf{Baseline vs. Accelerated Markov Chain Pipeline for Stationary Distribution Prediction.} Our method accelerates Step 1 by reparameterizing the integral bounds and treating dead time as a row permutation, enabling an efficient transition matrix construction without sequential computation. This results in a $1000 \times$ acceleration in the prediction of distorted stationary PDFs due to dead time.}
    \label{fig:main}
    \vspace{-15pt}
\end{figure*}

The main computational bottleneck in Markov-based modeling lies in the per-element evaluation of integrals when constructing the transition matrix. We develop an equivalent formulation that enables fully vectorized computation. The key ideas are: (i) replace each integral with a difference of cumulative flux functions, and (ii) treat dead time as a circular permutation of rows in the matrix. This allows the transition matrix to be built in parallel without explicit loops. Figure~\ref{fig:main} summarizes the Markov chain pipeline for stationary distribution prediction and highlights where the acceleration takes place.

\subsection{Transition Kernel}

According to~\cite[Proposition~1]{Rapp_2019_Dead}, for two consecutive registrations $x_k \to x_{k+1}$, the transition PDF is
\begin{equation}
    \label{eq:transition_pdf}
    f_{X_{k+1}|X_k}(x_{k+1}|x_k) = \frac{\lambda(x_{k+1})}{1 - \exp(-\Lambda)} \exp\left[- \int_{a}^{b} \Tilde{\lambda}(\tau) \, d\tau\right],
\end{equation}
where we define $a \coloneq x_k+x_d$, $b \coloneq \left\lceil\frac{x_k+x_d-x_{k+1}}{t_r}\right\rceil t_r + x_{k+1}$, and $\Tilde{\lambda}(t)$ is a periodic replica of $\lambda(t)$. $x_d$ denotes the relative dead time, i.e. $x_d \coloneq t_d \bmod t_r$, and $\lceil \cdot \rceil$ is the ceiling function.

To discretize the PDF, we define a discretization operator $\mathcal{D}(\cdot) : [0, t_r) \to \mathcal{S}$ as $\mathcal{D}(t) \coloneq \left(\left\lceil \frac{t}{\Delta} \right\rceil - \frac{1}{2}\right)\Delta$, moving continuous times to bin centers. Then, the transition probability is
\begin{equation}
    \label{eq:transition_pmf}
    P_{ij} = \Pr(x_{k+1} = s_j \, | \, x_k = s_i ) \propto \lambda(s_j)\,E_{ij},
\end{equation}
where $E_{ij} = \exp\!\Big[- \int_{\mathcal{D}(a)}^{\mathcal{D}(b)} \Tilde{\lambda}(\tau)\, d\tau \Big]$ is the exponent term with limits depending on $s_i$, $s_j$, and $x_d$, whose coupling forces per-element integrals, impeding straightforward parallelization.

\subsection{Equivalent Exponent Term}

A key observation is that the lower and upper integral limits represent the reactivation point after dead time and the next registration, respectively. The complex form of $b$ enforces it to be larger than $a$, and ensures the positivity of the integral. Alternatively, we keep $x_{k+1}$ at the upper bound and map the reactivation point to the same relative coordinate
\[
x_k' = (x_k + x_d) \bmod t_r \bydef x_k \oplus x_d,
\]
where $\oplus$ denotes circular addition under $t_r$. Thus, the integral region is always between $x_k'$ and $x_{k+1}$ (possibly wrapping around).

\boxedthm{
\begin{theorem}
\label{thm:equi_exponent}
We propose an equivalent representation of the exponent term in Eq.~\ref{eq:transition_pmf}:
\begin{equation}
    \label{eq:Eij_ori}
    E_{ij} = \exp\!\Big[-\Lambda \cdot \ind{\mathcal{D}(x_k^{\prime}) > s_j} - F(s_j) + F(\mathcal{D}(x_k^{\prime})) \Big],
\end{equation}
where $F(t) \coloneq \int_0^t \lambda(\tau)\, d\tau$ is the cumulative flux.
\end{theorem}
}

\noindent\textbf{Remark 1.} \emph{The proof follows by enumerating the four possible relationships between $x_k + x_d$ and $x_{k+1}$ and analyzing the exponent term in Eq.~\ref{eq:transition_pmf}, each of which yields the same cumulative-flux in Eq.~\ref{eq:Eij_ori}. This equivalent form reduces three coupled variables to two and compresses the periodic $\Tilde{\lambda}(t)$ to one cycle, serving as a key step toward a parallel matrix construction.}

\subsection{Separation of Dead Time}

With Theorem~\ref{thm:equi_exponent}, the remaining challenge is the term $\mathcal{D}(x_k^{\prime}) = \mathcal{D}(x_k \oplus x_d)$. This quantity is obtained by a “shift-then-quantize” operation, which couples shifting with discretization and becomes a bottleneck for vectorized matrix construction from $s_i$ to $s_j$. To overcome this, we introduce an alternative operation that separates the discretization from the dead time effect, making the structure amenable to efficient matrix construction.

\boxedthm{
\begin{theorem}
\label{thm:equi_operator}
The ``shift-then-quantize'' operation can be approximated by a ``quantize-then-shift'' operation
\[
\mathcal{P}(x_k) \bydef \mathcal{D}(x_k) \oplus \Big\lceil \tfrac{x_d}{\Delta} \Big\rceil \Delta = s_i \oplus \Big\lceil \tfrac{x_d}{\Delta} \Big\rceil \Delta.
\]
\end{theorem}
}

\noindent\textbf{Remark 2.} \emph{The error vanishes as bin resolution increases. This interpretation reveals that dead time corresponds to a row-wise circular shift of the transition matrix from state $s_i$ to $s_j$.}

\noindent Therefore, the exponent term in Eq.~\ref{eq:Eij_ori} becomes
\begin{equation}
    \label{eq:Eij_ours}
    E_{ij} = \exp\left[- \Lambda \cdot \ind{\mathcal{P}(x_k)  > s_j} - F(s_j) + F(\mathcal{P}(x_k)) \right].
\end{equation}

\subsection{Parallel Matrix Construction}

\begin{figure*}[t]
    \centering
    \includegraphics[width=0.97\linewidth]{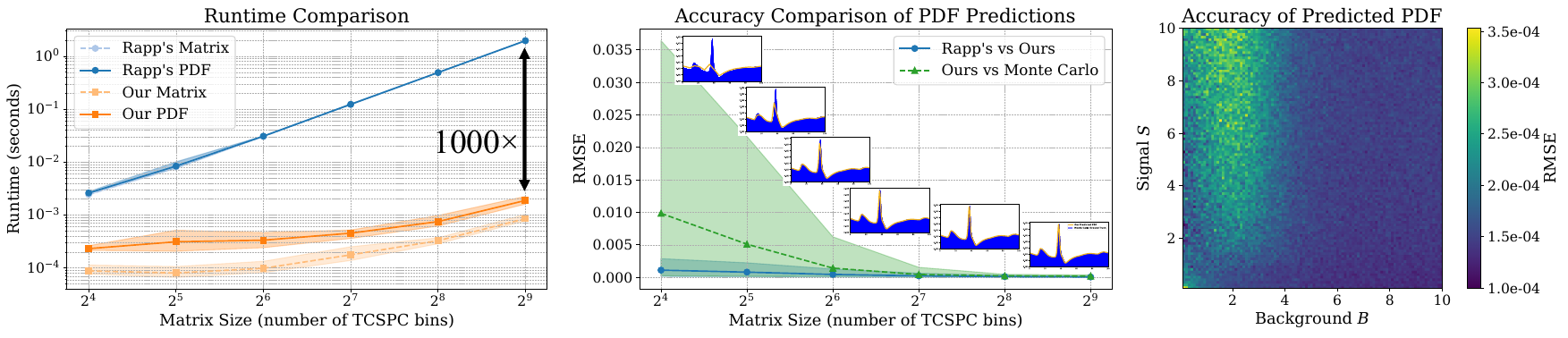}
    
    \caption{\textbf{Runtime and Accuracy of Accelerated Stationary PDF Prediction.}
    (a) Runtime versus matrix size for Rapp’s method and ours (evaluated at $S,B \in \{0.1,3,6,9\}$).
    (b) Prediction errors relative to Rapp’s method and to the Monte Carlo reference (same $S,B$ set).
    Shaded bands show min–max; markers denote means.
    (c) RMSE relative to the Monte Carlo reference over a grid with $S,B \in [0,10]$.}
    \label{fig:method_evaluation}
    \vspace{-10pt}
\end{figure*}

With $E_{ij}$ specified, we assemble the matrix $\mE$ in a vectorized manner. The exponent matrix $\mE$ is built from a vector of current states $\mathbf{x_k} = \mathbf{s} = (s_1, s_2, \ldots, s_{n_b})^{\top}$ and a vector of next states $\mathbf{x_{k+1}} = \mathbf{s}$. Without dead time, $\mathcal{P}(\mathbf{x_k}) = \mathbf{x_k} = \mathbf{s}$, and the exponent matrix is
\begin{equation}
    \mE_{\text{base}} = \exp^{\odot}\!\big(-\Lambda \mL - \mD\big),
\end{equation}
where $\mL \coloneq \ind{\mathbf{s} \mathbf{1}^\top > \mathbf{1} \mathbf{s^\top}}$ is a strictly lower-triangular indicator matrix ($\mathbf{1}$ is a column vector with all ones), $\mD \coloneq \mathbf{1}\mathbf{F}^\top(\mathbf{s}) - \mathbf{F}(\mathbf{s}) \mathbf{1}^\top$ encodes pairwise cumulative differences, and $\odot$ is the Hadamard product representing element-wise calculation.

With dead time, $\mE = \mJ_l \mE_{\text{base}}$ where $\mJ_l$ is a permutation matrix that rolls matrix rows by $l=\lceil x_d/\Delta \rceil$, according to Theorem~\ref{thm:equi_operator}. The transition kernel is then
\begin{equation}
    \mP = (\mathbf{1}\,\vlambda^\top) \odot \mE,
\end{equation}
with $\vlambda_j = \lambda(s_j)$. After row normalization, this yields the transition matrix $\Tilde{\mP}$ whose stationary eigenvector gives the registration PDF.

\subsection{Runtime and Accuracy Evaluation}

\begin{figure*}[t]
    \centering
    \includegraphics[width=0.97\linewidth]{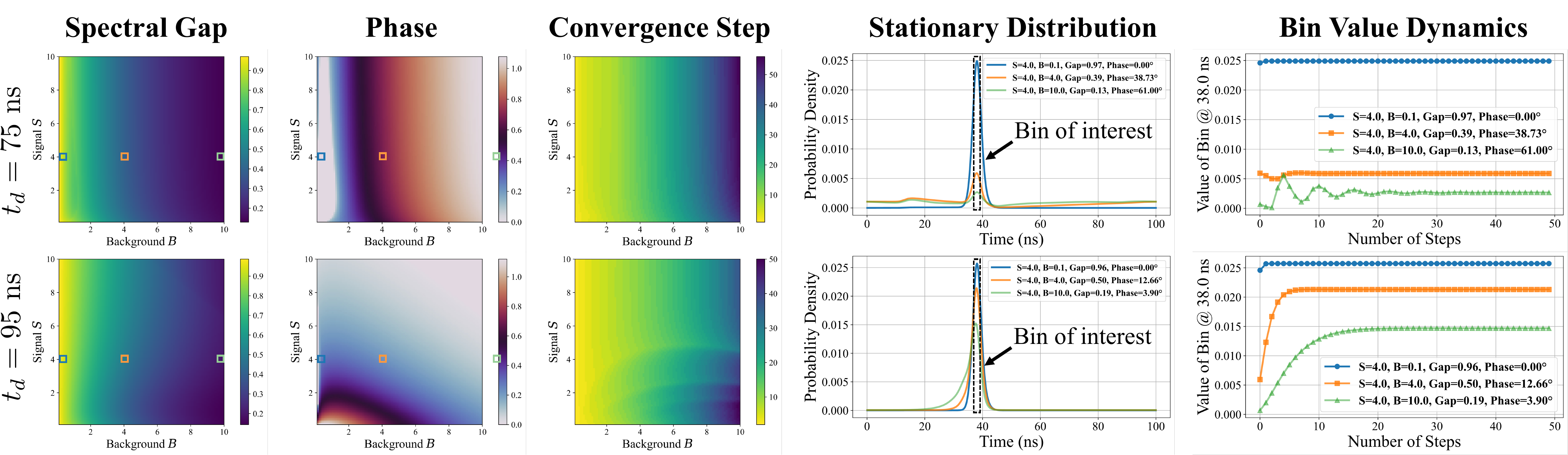}
    \caption{\textbf{Spectral Convergence Analysis.}
    (1) Spectral gap $1-|\lambda_2|$ across $(S,B)$ grids. 
    (2) Phases of $\lambda_2$ showing oscillatory patterns under different dead times $t_d$. 
    (3) Empirical mixing steps, consistent with the spectral gap. 
    (4) Convergence dynamics of a selected histogram bin, illustrating amplitude decay governed by the gap and oscillation frequency governed by the phase.}
    \label{fig:convergence_analysis}
    \vspace{-15pt}
\end{figure*}

We evaluate our accelerated Markov modeling method in terms of runtime and accuracy against the baseline approach of Rapp~\cite{Rapp_2019_Dead}, and also compare our method with the Monte Carlo ground truth.

\textbf{Ours vs. Rapp's Method:} 
Figure~\ref{fig:method_evaluation}(a) shows that the runtime of Rapp's element-wise construction grows rapidly with matrix size, whereas our parallelized scheme remains stable. Meanwhile, as shown in Fig.~\ref{fig:method_evaluation}(b), our predictions are numerically consistent with Rapp's, with discrepancies diminishing as the bin resolution increases. Overall, our method achieves $1000 \times$ speedup without loss of accuracy.

\textbf{Ours vs. Monte Carlo:} 
The Markov chain model is a discretized approximation. To quantify modeling fidelity, we compare our predictions against Monte Carlo estimates obtained by averaging $100$ empirical histograms, each from $50{,}000$ laser cycles at a fine resolution ($2^{10}$ bins). Figure~\ref{fig:method_evaluation}(b) shows that once the number of bins exceeds $2^7$, the Markov-chain-based PDFs closely match the Monte Carlo reference.

\textbf{Accuracy across $(S,B)$:} 
To further assess robustness, we evaluate prediction accuracy across a grid of signal and background levels $S,B \in [0,10]$. With $2^8$ bins, each Monte Carlo reference is the empirical average of $25$ runs of $50{,}000$ cycles. Figure~\ref{fig:method_evaluation}(c) confirms that our predictions remain quantitatively consistent with Monte Carlo across all $(S,B)$ levels.

Our results demonstrate that our method not only accelerates computation by orders of magnitude but also preserves accuracy.

\section{Spectral Convergence Analysis}
\label{sec:convergence}

With the accelerated construction of the transition matrix $\Tilde{P}$, we study the spectral convergence of the Markov chain — a problem that was previously computationally intractable and largely unexplored. While there is a consensus that the convergence is determined by the magnitude of the second largest eigenvalue $|\lambda_2|$ of $\Tilde{P}$~\cite{Rapp_2019_Dead}, quantitative results on mixing times and the role of eigenvalue phases under different $(S,B,t_d)$ have remained open.

\subsection{Spectral Gap and Mixing Time}

Figure~\ref{fig:convergence_analysis} (first column) shows the spectral gap $1-|\lambda_2|$ across $(S,B)$ grids. The gap decreases as background $B$ increases, indicating slower convergence under high ambient light. The third column reports the number of steps $n$ such that $\Tilde{P}^n$ approaches the stationary distribution, which aligns closely with the spectral gap and validates the theoretical link between $|\lambda_2|$ and mixing time.

\subsection{Eigenvalue Phase and Oscillatory Dynamics}

Interestingly, as shown in Fig.~\ref{fig:convergence_analysis} (second column), phase patterns vary with $t_d$: for $t_d=75$, the phase grows with $B$, while for $t_d=95$ a large region exhibits near-zero phase, resembling signal gating.

To illustrate the oscillatory effects of phase, we select three $(S,B)$ pairs and examine the temporal evolution of one representative histogram bin in the stationary PDFs. The results show that the spectral gap dictates the convergence rate in amplitude, while the phase determines the oscillation frequency. This demonstrates that both eigenvalue magnitude and phase are necessary to fully characterize convergence behavior in asynchronous SP-LiDAR.

\section{Conclusion}
We presented an accelerated framework for modeling timestamp distributions in asynchronous SP-LiDAR via an efficient construction of the Markov chain transition matrix. The formulation reduces the integral evaluation to cumulative flux differences and interprets dead time as a permutation, enabling parallel matrix assembly. This acceleration not only provides orders-of-magnitude speedup over existing methods but also makes possible new analyses of stationary distributions and spectral convergence that were previously out of reach. These results open the door to principled characterization of detector dynamics under high-flux conditions and support the design of faster, more accurate SP-LiDAR systems.




\bibliographystyle{IEEEbib}
\bibliography{main}

\end{document}